\documentclass[twocolumn, pra, showpacs, showkeys]{revtex4}
\usepackage{graphicx}
\usepackage{dcolumn}
\usepackage{bm}

\begin{document}

\preprint{APS/123-QED}

\title{Bose-Einstein condensation of trapped interacting spin-1 atoms}
\author{Wenxian Zhang,$^1$ Su Yi,$^2$ and L. You$^1$}
\affiliation{$^1$School of Physics, Georgia Institute of
Technology, Atlanta, Georgia 30332-0430, USA\\
$^2$Department of Physics and Astronomy and Rice Quantum
Institute, Rice University, Houston, Texas 77251-1892, USA}
\date{ 14 June 2004 }

\begin{abstract}
We investigate Bose-Einstein condensation of trapped spin-1 atoms
with ferromagnetic or antiferromagnetic two-body contact
interactions. We adopt the mean field theory and develop a
Hartree-Fock-Popov type approximation in terms of a semiclassical
two-fluid model. For antiferromagnetic interactions, our study
reveals double condensations as atoms in the $|m_F=0\rangle$ state
never seem to condense under the constraints of both the
conservation of total atom number $N$ and magnetization $M$. For
ferromagnetic interactions, however, triple condensations can
occur. Our results can be conveniently understood in terms of the
interplay of three factors: (anti) ferromagnetic atom-atom
interactions, $M$ conservation, and the miscibilities between
and among different condensed components.
\end{abstract}



\maketitle
\narrowtext

\section{Introduction}
Research in dilute atomic quantum gases remains one of the most
vibrant areas in physics almost ten years after the landmark
discovery of atomic Bose-Einstein condensation (BEC).
Increasingly, new experiments are revealing the rich possibilities
afforded by the internal electronic state structures of an atom,
e.g., the generation of atom entanglement in the Mott phase through
controlled collisions from the relative displacement of the
optical lattice potentials of the respective atomic internal
states \cite{becent} and the recent observation of condensation of
fermionic atom pairs \cite{jin}.

It has been known for a long time that the spinor degrees of
freedom of an atom becomes accessible if a far-off-resonant
optical trap is used to provide equal confinement for all Zeeman
states, instead of the more widely used magnetic traps for spin
polarized atoms \cite{Ho98, Law, Ohmi98,simkin}. Several earlier
experiments have produced fascinating observations of spinor
condensates, a superfluid with internal degrees of freedom, e.g.,
with $^{23}$Na atoms in $F=1$ \cite{Ketterle98} and $F=2$
\cite{Ketterle03} and $^{87}$Rb atoms in $F=1$ \cite{Chapman01,
Erhard04} and $F=2$ \cite{Chapman03, Schmaljohann04}, spin domains
and interdomain tunnelling \cite{Miesner99, Kurn99}, as well as
the generation of coreless vortex states \cite{Yip99, Mizushima02,
Martikainen, Ketterle03Vortex}. These properties exist only
because of the spinor nature of the condensate order parameter,
and thus are generally not expected to occur in a magnetically trapped
condensate.

Despite these and other related successes with spinor condensates,
our knowledge remains limited regarding the condensation
thermodynamics of spin-1 atoms. In a sense, the spin-1 condensate
constitutes a type of quantum fluid unfamiliar to many of us.
On the experimental side, it remains a significant challenge to
produce a spinor condensate, as evidenced by the disproportionally
small numbers of spinor BEC experiments in operation. In this
article, we reconsider the topic of the condensation thermodynamics
for a system of trapped spin-1 atoms. Of particular interest to us
is the question of the so-called double condensations for a spin-1
system constrained by two global conservations \cite{Machida}.
Using the Bogoliubov-Popov approximation, Isoshima {\it et al.}
first investigated the thermodynamics of the BEC phase transition
for a spin-1 gas \cite{Machida}. Huang {\it et al.} studied
analytically the effect of a magnetic field on the transition
temperature \cite{Tsai}. While an attempt to find the
zero-magnetic-field phase diagram was made through numerical
simulations in Ref. \cite{Machida}, there still exist several
question marks to the overall picture of BEC for a spin-1 Bose
gas, especially for ferromagnetically interacting atoms such as
$^{87}$Rb. Limited by the computation procedure within the
Bogoliubov-Popov approximation, only a few data points were made
available in the earlier studies by Isoshima {\it et al.}
\cite{Machida}. The lack of focused experimental efforts also
indirectly discouraged a detailed investigation of this problem
until now.

In this paper we systematically investigate the phase diagram of a
spin-1 Bose gas for both ferromagnetic and antiferromagnetic interactions.
Instead of the Bogoliubov approach, we will adopt the Hartree-Fock-Popov
approximation and employ a semiclassical approximation to the noncondensed
atoms within the mean field theory. We will also enforce the thermodynamics
for a finite trapped system with a fixed
total atom number $N$ and a total magnetization $M$.
Recent studies have significantly verified the accuracy of this approximation
when applied to similar systems \cite{Gerbier04}.
As we will illustrate in this work
our results indicate that double condensations will occur for
a spin-1 gas with antiferromagnetic interactions, while
triple condensations are more likely for ferromagnetic interactions.

This paper is organized as follows. In Sec. II, we review
the additional features of a BEC for an ideal gas of spin-1 atoms.
This is followed by the discussion of an interacting spin-1 gas
in Sec. III and a brief sketch of the Hartree-Fock-Popov theory used
for our investigation. We outline the detailed numerical algorithm
used to solve the coupled two-fluid model quantum gas at different
temperatures in Sec. IV and present the results of our study
in Sec. V. We conclude with some discussions and remarks in Sec. VI.

\section{BEC of an ideal gas of spin-1 atoms}
In this section, we briefly review the phenomenon of a BEC for a trapped
noninteracting gas of spin-1 atoms following the pioneering study of
Isoshima {\it et al.} \cite{Machida}.
At thermal equilibrium, we adopt the standard
Bose-Einstein distribution, and treat the spinor degree of
freedom as degenerate internal states in the absence of an external magnetic field.
The average number of atoms at each single
atom state of an energy $\varepsilon_j$ for the component
$|F=1,m_F=i\rangle$, $i=+1,0$, and $-1$ (hereafter $|i\rangle$),
is then conveniently given by
\begin{eqnarray}
N_{i,j}= {{\sf z_i}e^{-\beta \varepsilon_j}\over 1-{\sf z}_i e^{-\beta \varepsilon_j}},
\label{bed}
\end{eqnarray}
with $\beta=1/(k_BT)$ at temperature $T$.
$k_B$ is the Boltzmann constant.
The fugacity ${\sf z}_i$ can be expressed in terms of the chemical
potential for the $i$th component $\mu_i$ as ${\sf z}_i=\exp(\beta\mu_i)$.
In the thermodynamic limit, one can follow the usual approach by
making a semiclassical approximation for a continuous description of the
single particle density of states, and treating the ground state population
separately as it can become macroscopic due to Bose-Einstein condensation.
The total number of atoms for a given internal state
in all motional excited states of the trap is thus found to be
\begin{eqnarray}
N_i^T&=&\sum_{j=1}^\infty N_{i,j} = \left({k_BT \over \hbar \omega }\right)^3
g_3({\sf z}_i),
\end{eqnarray}
where we have assumed atoms are confined in a spherical
harmonic trap with a frequency $\omega$ independent of the
atomic internal state $|i\rangle$.
$g_\xi(x)=\sum_{n=1}^\infty
(x^n/n^\xi)$ is the standard Bose function \cite {Pritchard87,Ketterle96}.
We note that the conservations of the
total number of atoms $N=N_++N_0+N_-$ and total magnetization
$M=N_+-N_-$ lead to the chemical potentials for different spin
components expressible as $\mu_{\pm}=\mu\pm\eta$ and $\mu_0=\mu$.
These identities remain valid in the presence of atom-atom interactions.
$\mu$ and $\eta$ are effectively
independent Lagrange multipliers used
to guarantee the conservation of $N$ and $M$, respectively.
Taking the single atom trapped ground state to be zero energy,
the Bose distribution (\ref{bed})
shows that $\mu_i$ is negative at high temperatures and reaches
zero when the spin component $|i\rangle$ condenses.
$\eta$ is positive (negative) for a positive $M$ (negative),
which acts as a fictitious applied magnetic field physically.

As was first pointed out in Ref. \cite{Machida}, there exists an interesting
double condensation phenomenon for a spin-1 gas because of the presence of
$M$ conservation.
When the temperature is lowered,
the $|+\rangle$ component first condenses for a system with a positive $M$
because its phase space density is largest, reflecting the fact that
$N_+$ is the largest component population. Thus we first arrive at $\mu_+=0$.
This consideration leads to the critical temperature $T_1$ governed
by the following equations:
\begin{eqnarray}
N &=& \left({k_BT_1 \over \hbar \omega }\right)^3 \left[g_3(1) +g_3(e^{\beta\mu})
+g_3(e^{2\beta\mu})\right],\label{eqT1a}\\
M &=& \left({k_BT_1 \over \hbar \omega }\right)^3\left [g_3(1)-g_3(e^{2\beta\mu})\right].
\label{eqT1b}
\end{eqnarray}
On further lowering of the temperature,
however, the remaining two components $|0\rangle$ and
$|-\rangle$ condense simultaneously, rather than sequentially
with the less populated component of the two condensing last.
This is precisely due to the conservation identities as
discussed before. The relationships $\mu_{\pm}=\mu\pm\eta$ and $\mu_0=\mu$
lead to a mathematical certainty:
when $\mu_+$ is zero,
if either $\mu_0$ or $\mu_-$ becomes zero, both must be zero.
At this second critical temperature $T_2$, both $\mu_0=0$ and $\mu_-=0$,
which imply
that the $|0\rangle$ and $|-\rangle$ components condense simultaneously.
This second condensation where all three components condense, occurs at
the temperature $T_2$ of
\begin{eqnarray}
T_2 &=& {\hbar\omega\over k_B}\left[{N-M\over 3g_3(1)}\right]^{1/3}.
\end{eqnarray}

\begin{figure}
\includegraphics[width=3.25in]{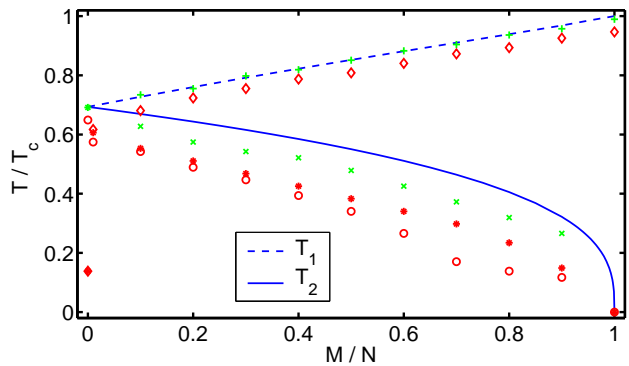}
\caption{BEC for a gas of spin-1 atoms with $M>0$ ($M<0$).
For noninteracting atoms,
the $|+\rangle$ ($|-\rangle$) component condenses first at $T_1$ (dashed line)
while the $|0\rangle$ and $|-\rangle$ ($|+\rangle$) components condense
simultaneously at $T_2$ (solid line).
For $^{23}$Na atoms with antiferromagnetic interactions,
double condensations persist according to our theoretical study.
The $|+\rangle$ component (denoted by +) condenses first, which is then
followed by the condensation of the
$|-\rangle$ component (denoted by $\times$). The $|0\rangle$ component
is unpopulated in the low temperature limit.
For $^{87}$Rb atoms with ferromagnetic interactions, our study reveals
the potential for triple condensations.
First, the $|+\rangle$ component condenses (denoted by
$\diamond$), which is then followed by the second condensation for the $|-\rangle$
component (denoted by $*$), and finally the third condensation for
the $|0\rangle$ component occurs (denoted by $\circ$).}
\label{pc}
\end{figure}

In Fig. \ref{pc}, we illustrate the $M$ dependence of the
double condensations for an ideal Bose gas of spin-1 atoms.
$T_c=[N/g_3(1)]^{1/3}(\hbar\omega/k_B)\approx 0.94N^{1/3}\hbar\omega/k_B$
is the condensation temperature for $M=N$, i.e., for a single component
gas with all atoms polarized in state $|+\rangle$.

\section{BEC of an interacting gas of spin-1 atoms}

\subsection{Formulation}
Our model system of the interacting spin-1 atoms
is described by the following
Hamiltonian in second quantized form
\begin{eqnarray}
H&=&\int d\vec r\left\{ \psi_i^\dag\left[-{\hbar^2\over 2m}\nabla^2+V_{\rm
ext}(\vec r)\right]
\psi_i+{c_0\over 2}\psi_i^\dag\psi_j^\dag\psi_j\psi_i\right.\nonumber\\
&&\left.+{c_2\over 2}\psi_i^\dag\left(F_\alpha\right)_{ij}\psi_j
\psi_k^\dag\left(F_\alpha\right)_{kl}\psi_l\right\},
\label{ham}
\end{eqnarray}
where $\psi_i(\vec r)$ [$\psi_i^\dag(\vec r)$] is the quantum field
for annihilating an atom in state $|i\rangle$ at $\vec r$, and
$i,j,k,l=+,0,-$ and $\alpha=x,y,z$.
Repeated indices are assumed to be summed \cite{Ho98, Ohmi98}.
$F_{\alpha=x,y,z}$ are spin-1 matrices given by
\begin{eqnarray}
F_x&=&{1\over\sqrt{2}}\left(\begin{array}{ccc}0&1&0\\1&0&1\\0&1&0
\end{array}\right), \nonumber\\
F_y&=&{i\over\sqrt{2}}\left(\begin{array}{ccc}0&-1&0\\1&0&-1\\0&1&0
\end{array}\right),\nonumber\\
F_z&=&\left(\begin{array}{ccc}1&0&0\\0&0&0\\0&0&-1\end{array}\right),
\nonumber
\end{eqnarray}
with the quantization axis taken along the z-axis direction.
It is easy to check that both the total number of atoms
and the total magnetization
\begin{eqnarray}
N&=&\int d\vec r (|\psi_+|^2+|\psi_0|^2+|\psi_-|^2), \nonumber\\
M&=&\int d\vec r (|\psi_+|^2-|\psi_-|^2)=\int d\vec r \psi_i^\dag
(F_z)_{ij} \psi_j,\nonumber
\end{eqnarray}
commute with the above Hamiltonian (\ref{ham}), and are thus constants of motion.
To study the minimal energy ground state, we therefore introduce two
Lagrange multipliers $\mu$ and $\eta$, to fix the total atom number and
magnetization of the system in our numerical minimization. It turns out that
$\mu$ is in fact the chemical potential of the system and
$\eta$ is an effective magnetic field. The Gibbs free energy is then given by
\begin{eqnarray}
G&=&H-\mu N- \eta M\nonumber\\
&=&\int d\vec r \left\{\psi_i^\dag({\cal L}_{ij}-\eta (F_z)_{ij}) \psi_j
+{c_0\over 2}\psi_i^\dag\psi_j^\dag\psi_j\psi_i\right.\nonumber\\
&&+\left.{c_2\over 2}\psi_i^\dag\left(F_\alpha\right)_{ij}\psi_j
\psi_k^\dag\left(F_\alpha\right)_{kl}\psi_l\right\},
\end{eqnarray}
where
\begin{eqnarray}
{\cal L}_{ij}&=&\left[-{\hbar^2\over 2m}\nabla^2-\mu+V_{\rm
ext}(\vec r)\right]\delta_{ij}. \nonumber
\end{eqnarray}
The atomic interactions are conveniently parametrized
through the two $s$-wave scattering lengths $a_0$ and $a_2$ between two spin-1 atoms
\cite{Ho98,Ohmi98,Law}
\begin{eqnarray}
c_0&=&{4\pi\hbar^2\over m}\left({a_0+2a_2\over 3}\right), \nonumber\\
c_2&=&{4\pi\hbar^2\over m}\left({a_2-a_0\over 3}\right).\nonumber
\end{eqnarray}
In this study, we attempt to find
the mean field ground state of our system, which
corresponds to the state with the lowest Gibbs free energy.

\subsection{Hartree-Fock-Popov theory and the two-fluid model approximations}
The field operator $\psi(\vec r, t)$ evolves in the Heisenberg picture according to
\begin{eqnarray}
i\hbar \frac {\partial}{\partial t}\psi(\vec r, t) &=& \left[
\psi, G \right]. \nonumber
\end{eqnarray}
For the system of a spin-1 Bose gas as considered here,
the above equation becomes
\begin{widetext}
\begin{eqnarray}
i\hbar \frac {\partial}{\partial t}\psi_+(\vec r, t) &=& {\cal
L}_{++}\psi_+ - \eta \psi_+ +
c_0\sum_j\left(\psi_j^\dag\psi_j\right)\psi_+
 + c_2\left[
\left(\psi_+^\dag\psi_++\psi_0^\dag\psi_0-\psi_-^\dag\psi_-\right)
\psi_+ + \psi_-^\dag\psi_0\psi_0\right], \nonumber\\
i\hbar \frac {\partial}{\partial t}\psi_0(\vec r, t) &=& {\cal
L}_{00}\psi_0 + c_0\sum_j\left(\psi_j^\dag\psi_j\right)\psi_0
 + c_2\left[ \left(\psi_+^\dag\psi_+ + \psi_-^\dag\psi_-\right)
\psi_0 + 2\psi_0^\dag\psi_+\psi_-\right],\\
i\hbar \frac {\partial}{\partial t}\psi_-(\vec r, t) &=& {\cal
L}_{--}\psi_- + \eta \psi_- +
c_0\sum_j\left(\psi_j^\dag\psi_j\right)\psi_-
 + c_2\left[
\left(\psi_-^\dag\psi_-+\psi_0^\dag\psi_0-\psi_+^\dag\psi_+\right) \psi_- +
\psi_+^\dag\psi_0\psi_0\right].\nonumber
\end{eqnarray}
Following the standard mean field theory procedure, i.e. taking
$\psi=\phi+\delta\psi$ with $\phi=\langle\psi\rangle$, after tedious
manipulations and calculations, we obtain
a set of coupled Gross-Pitaevskii equations for the
superfluid components
including their interactions with the noncondensed atoms as
\begin{eqnarray}
i\hbar {\partial \over \partial t}\phi_+ &=& \left[-{\hbar^2\over
2m}\nabla^2+V_{\rm ext} - \mu -\eta+c_0(n+n_+^T)
+c_2(n_++n_0-n_-+n_+^T)\right]\phi_++c_2\phi_0^2\phi_-^*,\nonumber\\
i\hbar {\partial \over \partial t}\phi_0 &=& \left[-{\hbar^2\over
2m}\nabla^2+V_{\rm ext} - \mu +c_0(n+n_0^T)
+c_2(n_++n_-)\right]\phi_0+2c_2\phi_+\phi_-\phi_0^*,\label{gpe}\\
i\hbar {\partial \over \partial t}\phi_- &=& \left[-{\hbar^2\over
2m}\nabla^2+V_{\rm ext} - \mu +\eta +c_0(n+n_-^T)
+c_2(n_-+n_0-n_++n_-^T)\right]\phi_-+c_2\phi_0^2\phi_+^*,\nonumber
\end{eqnarray}
and equations for $\delta\psi_i$,
\begin{eqnarray}
i\hbar \frac {\partial}{\partial t}\delta\psi_+(\vec r, t) &=&
\left[-{\hbar^2\over 2m}\nabla^2 + V_{\rm ext} - \mu - \eta +
c_0(n+n_+) + c_2(2n_++n_0-n_-)\right]\delta\psi_+, \nonumber\\
i\hbar \frac {\partial}{\partial t}\delta\psi_0(\vec r, t) &=&
\left[-{\hbar^2\over 2m}\nabla^2 + V_{\rm ext} - \mu +
c_0(n+n_0) + c_2(n_++n_-)\right]\delta\psi_0,\label{exe}\\
i\hbar \frac {\partial}{\partial t}\delta\psi_-(\vec r, t) &=&
\left[-{\hbar^2\over 2m}\nabla^2 + V_{\rm ext} - \mu + \eta +
c_0(n+n_-) + c_2(2n_-+n_0-n_+)\right]\delta\psi_-,\nonumber
\end{eqnarray}
\end{widetext}
where $n=\sum_i n_i=\sum_i (|\phi_i|^2+n_i^T)$ is the
total density of the atomic gas, with $n_i^T=\langle\delta\psi^\dag_i \delta\psi_i\rangle$
the normal (noncondensed) gas density of the $i$th component.
Instead of the Hartree-Fock-Bogoliubov (HFB) approximation
as employed by Isoshima {\it et al.} \cite{Machida}, we have used
the Hartree-Fock-Popov (HFP) approximation to obtain the above
equations. Within the HFP approximation, we neglect
terms proportional to the anomalous noncondensate density
$\langle\delta\psi_i\delta\psi_j\rangle$ as well as their
complex conjugates. We have also
neglected terms proportional to
$\langle\delta\psi^\dag_i\delta\psi_j\rangle$ for $i\ne j$,
similar to the random phase approximation.
A more detailed formal discussion of the HFP theory can be found in
Refs. \cite{Str, Liu}, and for the calculation of the
phase diagram of Bose-Einstein condensation, it is
an excellent approximation as confirmed recently in a set of
detailed comparisons with experiments \cite{Gerbier04}.
In addition, as will become clear later, the HFP approximation
is also efficient from the numerical point of view, especially
near regions of temperatures close to (but below) the critical
temperature. The HFB approximation, on the other hand, is more
difficult to handle numerically \cite{Machida}. Although more rigorous
at very low temperatures,
the HFB approximation is expected to agree with the more
transparent HFP approximation at higher temperatures.
In deriving the equations for
$\delta\psi_i$, terms proportional to $\delta\psi_i^\dag$, $\delta\psi_j$,
and $\delta\psi_j^\dag$ for $j\ne i$ are also neglected.
This is equivalent to the neglect of the ``hole" component
in the HFB approximation, and is thus expected to have has a minor effect
except very close to the zero temperature.

In the HFP approximation we adopt here,
the normal fluid for noncondensed atoms is determined
through the semi-classical
approximation. We thus take $-i\hbar\nabla \rightarrow \vec p$,
and approximate its distribution by the standard Bose-Einstein
distribution in the phase space of $\{\vec p,\vec r\}$,
\begin{eqnarray}
n_i^T(\vec r)&=&\int {d\vec p\over (2\pi\hbar)^3}{1\over
e^{\varepsilon_i(\vec p,\vec r)/k_BT}-1},
\label{normal}
\end{eqnarray}
with the HFP single particle energy spectrum,
\begin{eqnarray}
\varepsilon_+(\vec p,\vec r)&=&{p^2\over 2m} + V_{\rm ext}
-\mu-\eta+c_0(n+n_+)\nonumber \\
&&+c_2\left(2n_++n_0-n_-\right),\nonumber\\
\varepsilon_0(\vec p,\vec r)&=&{p^2\over 2m}+V_{\rm
ext} -\mu+c_0(n+n_0)+c_2\left(n_++n_-\right),\nonumber\\
\varepsilon_-(\vec p,\vec r)&=&{p^2\over 2m}+V_{\rm ext}
- \mu+\eta+c_0(n+n_-)\nonumber \\
&&+c_2\left(2n_-+n_0-n_+\right),
\label{sptm}
\end{eqnarray}
which are obtained by substituting $\delta\psi_i(\vec
r,t)=\exp[-i\varepsilon_i(\vec p, \vec r) t/\hbar]u_i(\vec r)$ into Eqs.
(\ref{exe}) with $u_i(\vec r)$ the eigenfunction for the excitation of the
$i$th component.

Thus, we have formulated a coupled set of equations
for both the superfluid and the normal fluid; they are
Eq. (\ref{gpe}) for the condensed part and Eqs. (\ref{normal}) and (\ref{sptm})
for the noncondensed atoms. These are the basis for our numerical
investigations to be presented below.

\section{Numerical method}

In our numerical studies, we follow a standard procedure and
the following algorithm for the self-consistent solution of the coupled
equations (\ref{gpe}), (\ref{normal}), and (\ref{sptm})
as an extension of the single component gas studied earlier \cite{Str}.
Our algorithm is divided into the following steps:

\begin{itemize}
\item We find the condensate wave function
$\phi_i(\vec r)$ and the chemical potential $\mu$
for a set of fixed normal gas density $n_i^T(\vec r)$, by
propagating Eqs. (\ref{gpe}) in the imaginary time domain, as
described in Refs. \cite{Yi,NJP}.

\item We compute the updated energy spectrum and normal gas density
$n_i^T(\vec r)$ from Eqs. (\ref{normal}) and (\ref{sptm})
using the new condensate wave function and the chemical potential.

\item We normalize the total number of atoms to $N$ and adjust $\eta$
appropriately \cite{Yi}.

\item We repeat the above steps until final convergence is reached. The
convergence criterion is set to be that the condensate fraction $N^C(T)/N$ and the
magnetization fraction $M/N$ of successive iterations differ by
less than $10^{-11}$ for most temperatures and less than $10^{-5}$
near the phase transition temperature region.
\end{itemize}

At temperatures higher than the first BEC transition point, the above
procedure converges rather quickly as the superfluid component
$\phi_i$ is none existent. In this case, we only need to solve
Eqs. (\ref{normal}) and (\ref{sptm}) self-consistently
by adjusting $\mu$ and $\eta$.

\section{Results and discussion}

In this study, we focus on the illustration of our theory for atoms
inside a spherically symmetric harmonic trap
\begin{eqnarray}
V_{\rm ext}(\vec r) = {1\over 2}m\omega^2r^2.
\end{eqnarray}
We take $N=10^6$ and $\omega = (2\pi) 100$ Hz, and use the
spin-1 atom parameters for $^{23}$Na
and $^{87}$Rb atoms as given in Table \ref{tp}. Clearly
it is antiferromagnetic ($c_2>0$) for $^{23}$Na atoms
and ferromagnetic ($c_2<0$) for $^{87}$Rb atoms.

\begin{table}[h]
\caption{Atomic parameters for $^{23}$Na and $^{87}$Rb atoms \cite{para2, para3}.
$a_0$ and $a_2$ are in units of Bohr radius and $c_0$ and $c_2$ in units of
$10^{-12}$ Hz cm$^3$. }
\begin{tabular}{c|cccc}
\hline
&$a_0$&$a_2$&$c_0$&$c_2$\\
\hline
$^{23}$Na&50.0&55.0&15.587&0.4871\\
$^{87}$Rb&101.8&100.4&7.793&-0.0361\\
\hline
\end{tabular}
\label{tp}
\end{table}

\subsection{$^{23}$Na atoms with antiferromagnetic interactions}

\begin{figure}
\includegraphics[width=3.25in]{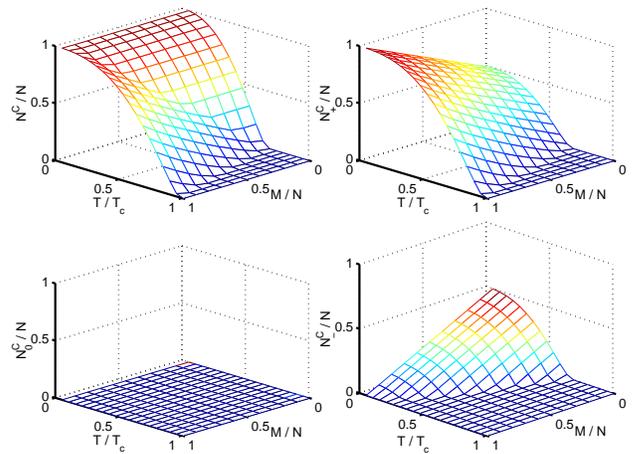}
\caption{Double condensations for a spin-1 gas of $^{23}$Na atoms. The upper left
panel shows the total condensed fraction vs temperature and total
magnetization. Similarly, the upper right one shows the fraction of condensed
$|+\rangle$ component, the lower left the condensed $|0\rangle$ component, and
the lower right the condensed $|-\rangle$ component.}
\label{Nap}
\end{figure}

\begin{figure}
\includegraphics[width=3.25in]{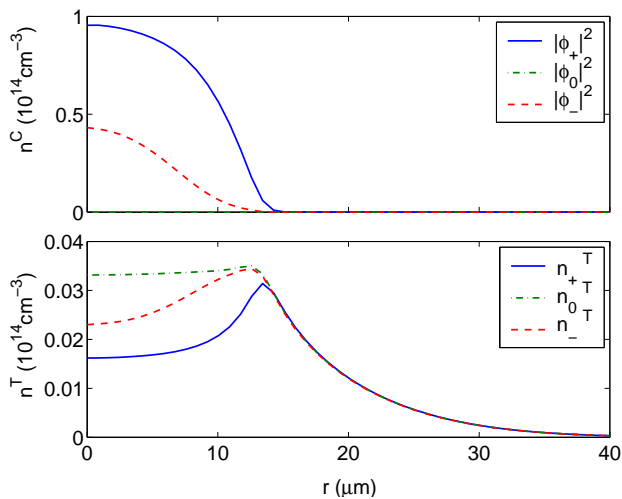}
\caption{Typical density distributions for different spin components of
a $^{23}$Na gas ($T/T_c=0.43, M/N=0.4$). The upper panel is for
the condensate and the lower one for the noncondensed atoms.}
\label{Nawf}
\end{figure}

The interaction between $^{23}$Na atoms is antiferromagnetic, i.e. $c_2>0$.
The phase diagram we obtain is shown in Fig. \ref{Nap}. It clearly
reveals the double
phase transitions: one for the $|+\rangle$ component and the other for the
$|-\rangle$ component. The $|0\rangle$ component of the condensate never shows up
because the antiferromagnetic interaction favors an antiparallel
alignment of the atomic spin, which is equivalent to a coherent
superposition of the $|+\rangle$ and $|-\rangle$ states as
explained in the discussion of order parameter symmetry
at zero temperature in Ref. \cite{Ho98}. Our mean field result is also consistent with
that of Isoshima {\it et al.} \cite{Machida}. Similar to the case of an ideal
gas, the transition
temperature of the $|+\rangle$ component increases monotonically with $M/N$
while that of the $|-\rangle$ component monotonically decreases.
When the
temperature decreases, the first condensed component is $|+\rangle$ because
$M>0$; the second condensed component is $|-\rangle$,
which condenses at temperatures when
$N_+^C+N_+^T-N_-^T>M$. Figure \ref{Nawf} shows typical density distributions
of different components for a $^{23}$Na gas.
We see that $|\phi_+|^2$ and
$|\phi_-|^2$ are always miscible \cite{KetterleNature, spin1} and distributed mostly
near the central region of the trap. We also note that $|\phi_0|^2$ is always
zero within this mean field study. All three components of the normal gas coexist. Both
$n_+^T$ and $n_-^T$ peak at the edge of the condensate because of the
shape of the net interaction potentials between
the condensate and the normal gas. $n_0^T$ is much flatter since
$|\phi_0|^2$ is zero.

We now comment on a particular feature related to the asymptotic behavior of
the spin-1 gas of $^{23}$Na atoms as $T\rightarrow 0$ for $M=0$. The full quantum
theory predicts a ground as a superfragmented Fock state
with atoms equally distributed among the three spin components $|N_+=N/3, N_0=N/3,
N_-=N/3\rangle$ \cite{Ho98, Law,Ho00, Ueda00}. Such a state would give rise to
a number fluctuation of order of $N^2$, and is impossible within the present
mean field treatment. The mean field ground state is known to be
$|N_+=N/2, N_0=0, N_-=N/2\rangle$ \cite{Ho00, Leggett01}, consistent with
our results.
In an actual experiment, it is most likely that the mean field ground state
is observed because the full quantum state is not stable against various
external sources of fluctuation or noises, e.g.
that of the unshielded magnetic field \cite{Ueda00}, a
small deviation of the total magnetization $M$ from zero \cite{Ho00}, or a
temperature being not exactly zero. The mean field ground state, on the other hand,
is more robust against these noise. In our numerical calculations, it is
really impractical to set the temperature
microscopically close to but above zero to probe the real ground state (for $M=0$).
We therefore enforced the ground state structure such that it
asymptotically approaches that of the mean field ground
state with a decrease of the temperature as shown in Fig. \ref{Nap}.
With this convention,
an related issue arises: the equivalence between states $|N_+=N/2, N_0=0, N_-=N/2\rangle$
and $|N_+=0, N_0=N, N_-=0\rangle$ at zero temperature as first pointed out by Ho
\cite{Ho98}. We note, however, that this equivalence is based on the assumption
of an environment perfectly free of magnetic fields. The presence of even a tiny
magnetic field, which is inevitable in the real world, will destroy this equivalence
and causing the real ground state to be $|N_+=N/2, N_0=0, N_-=N/2\rangle$,
a convention we chose as indicated in Fig. \ref{Nap}.

\subsection{$^{87}$Rb atoms with ferromagnetic interactions}

\begin{figure}
\includegraphics[width=3.25in]{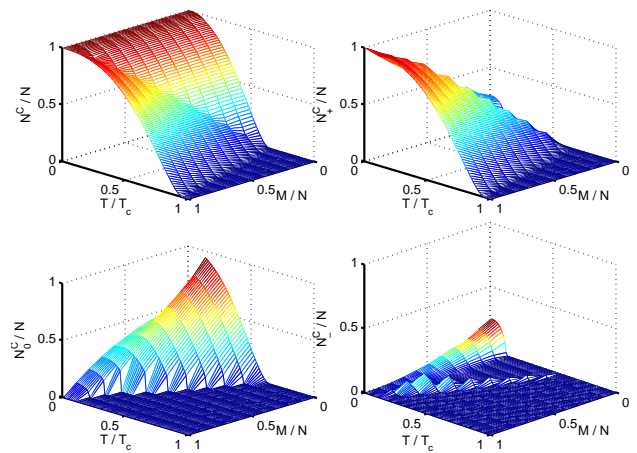}
\caption{The same as in Fig. \ref{Nap} but for a gas of $^{87}$Rb atoms.}
\label{Rbp}
\end{figure}

\begin{figure}
\includegraphics[width=3.25in]{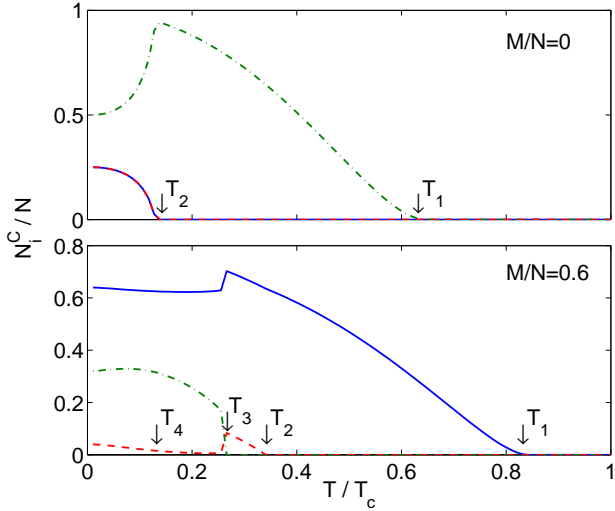}
\caption{Double condensations for $^{87}$Rb atoms when $M=0$ (the
upper panel) and triple condensations for $M/N=0.6$ (the lower panel).
The solid line denotes the fractional population of the condensed $|+\rangle$
component, the dot-dashed line denotes the $|0\rangle$ component,
and the dashed line denotes the $|-\rangle$ component.}
\label{Rbm}
\end{figure}

The phase diagrams for $^{87}$Rb atoms with ferromagnetic interactions ($c_2<0$)
are shown in Figs. \ref{Rbp} and \ref{Rbm}. Only a sparse set of points
weas made available in the early work of Isoshima {\it et al.} \cite{Machida}
because the numerical solution becomes far more difficult to converge in this case.
Based on our results, we see that when the temperature of the
system decreases, triple condensations occur in general.
When $M>0$, the first condensed component is
the $|+\rangle$ state (for $T<T_1$), the second one is
the $|-\rangle$ state (for $T<T_2$), and the last one is the
$|0\rangle$ state (for $T<T_3$). Our results show that the
$|+\rangle$ component first condenses at $T_1$ and its population
increases with decreasing temperature until $T_2$,
at which the $|+\rangle$ condensed component is a little more than
the total magnetization $M$. When temperature is lower than $T_2$,
the $|-\rangle$ component begins to condense as well. The two condensed
components in states $|+\rangle$ and $|-\rangle$ both increase with
decreasing temperature until the third critical temperature
$T_3$, at which the $|0\rangle$ component starts to condense.
Once the $|0\rangle$ component condenses, the $|-\rangle$ component
starts to decrease and becomes very close to zero, while
the $|+\rangle$ component is almost constant with respect
to further decreasing of the temperature. This trend continues
until the temperature is lower than $T_4$, when the condensed
$|0\rangle$ component starts to decrease with decreasing temperature while
the populations of the $|+\rangle$ and $|-\rangle$ condensed components
increase. For the
special case of $M=0$, on the other hand,
we observe only double condensations; the
$|0\rangle$ component condenses first, followed by the
simultaneous condensation of both the $|+\rangle$ and
$|-\rangle$ components. This is again due to the special symmetry
requirement that the $|+\rangle$ component must be the
same as the $|-\rangle$ component in order to keep $M=0$. We note that
in this case the fraction of condensed
$|0\rangle$ component can reach as high as $94\%$ at finite
temperatures, much higher than the $\sim 50\%$
at zero temperature.

\begin{figure}
\includegraphics[width=3.25in]{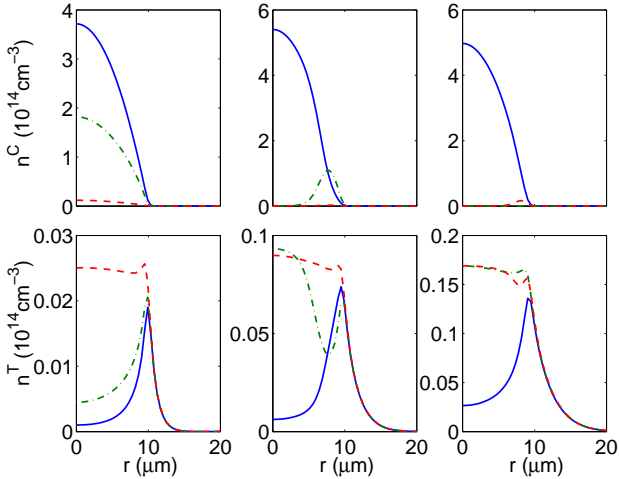}
\caption{Typical density distributions for a gas of $^{87}$Rb
atoms at $ M/N=0.6$. The left column is $T/T_c=0.11$, the middle
one is $0.21$, and the right one $0.32$. The notations are the same
as in Fig. \ref{Nawf}.}
\label{Rbwf}
\end{figure}

Figure \ref{Rbwf} displays typical density distributions for a gas of
$^{87}$Rb atoms at different temperatures for $M/N=0.6$.
The right column corresponds to $T\in
(T_3,T_2]$, where only the $|+\rangle$ and $|-\rangle$ components are
condensed and the $|-\rangle$ component is quite small and is
spatially located at the edge of the $|+\rangle$ component.
Quite generally, we note that with the condensation of a component,
its corresponding normal gas component would have a lower density.
For instance, the normal gas density of the $|+\rangle$ component is low in the trap
center where the $|+\rangle$ condensed component resides.
The middle column of Fig. \ref{Rbwf} is the typical density distribution
when $T\in (T_4,T_3]$.
The condensed $|+\rangle$ component stays at the center and is surrounded by
the $|0\rangle$ component. The condensed $|-\rangle$ component is too small to be
visible directly, but can be perceived from the
shallow well in its normal gas component, which
indicates that the condensed $|-\rangle$ component is not zero and
is located around the edge of the condensed $|+\rangle$ component.
The left column of Fig.
\ref{Rbwf} shows the density distributions when $T\in (0,T_4]$,
where all three
condensed components coexist near the center of the trap and are surrounded by
their normal gas components.

These results for $^{87}$Rb atoms can be understood in
terms of the interplay of three factors: ferromagnetic
atom-atom interaction ($c_2<0$), the
$M$ conservation, and the miscibility between and among
different components.
The ferromagnetic interaction favors the most populated state,
the $M$ conservation sets an upper limit on the
fraction of the condensed $|+\rangle$ component, and the
immiscibility between the condensed $|+\rangle$ and $|-\rangle$
component sets an upper limit on the total fraction of
the condensed $|+\rangle$ and $|-\rangle$ components.
For instance, in
the region $T\in(T_2, T_1]$, only the $|+\rangle$ component
condenses. The ferromagnetic interaction plays a dominant role
and thus more atoms condense into the $|+\rangle$ state with
decreasing temperature. In the region of $T\in(T_3, T_2]$, the $M$
conservation and the immiscibility begin to take their effect.
The $M$ conservation causes the increases to the condensed
$|+\rangle$ and $|-\rangle$ components to be almost identical,
while the immiscibility
makes the condensed $|-\rangle$ component stay outside the
condensed $|+\rangle$ component. The system becomes unstable with
the increase of the $|+\rangle$ and $|-\rangle$ components
because with more condensed $|+\rangle$ component,
the condensed $|-\rangle$ component must be pushed out further.
Near the third
critical temperature $T_3$, the condensed $|-\rangle$ component
suddenly decreases to almost zero and the condensed $|+\rangle$
component decreases to about $M$. Approximately, the total decreased amount
from the $|+\rangle$ and the $|-\rangle$ components
becomes the condensed $|0\rangle$ component. The system enters the
region $T\in(T_4, T_3]$ in which the condensed $|0\rangle$
component increases steadily with lowering temperature
because the condensed $|+\rangle$ and $|0\rangle$ components
are miscible. The condensed $|+\rangle$ component is almost
independent of the temperature because of the $M$ conservation.
With decreasing temperatures, more and more condensed $|0\rangle$
component finally suppresses the immiscibility between $|+\rangle$
and $|-\rangle$ component at $T_4$, all three condensed components
become miscible, and both the $|+\rangle$ and $|-\rangle$ components
increase to keep $M$ conserved while the $|0\rangle$ component
decreases.

\begin{figure}
\includegraphics[width=3.25in]{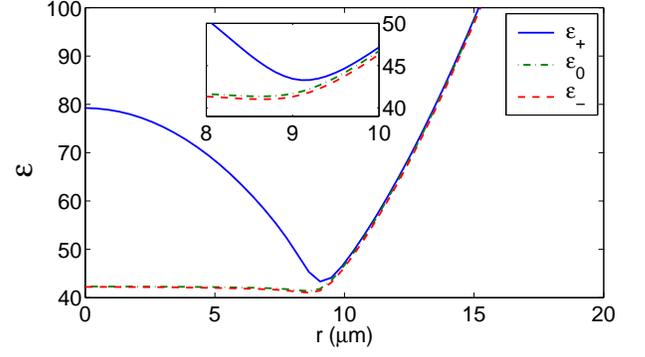}
\caption{The lowest excitation level for a gas of $^{87}$Rb atoms at $M/N=0.6$,
$T/T_c = 0.34$ (right before the condensation of the $|-\rangle$ component).
The inset shows the details of a zoomed-in plot near the minimum.}
\label{RbE}
\end{figure}

Figure \ref{RbE} shows the lowest excitation energy $\varepsilon = \varepsilon
(p=0)$ for the three components of a $^{87}$Rb gas at $M/N = 0.6$,
$T/T_c=0.34$ (right before the condensation of the $|-\rangle$ component).
We see that the energy for the
$|-\rangle$ component is lower than the corresponding ones for the other two
states and takes a minimum near the spatial location of
$r=9$ $\mu$m, which is at the edge of the condensed $|+\rangle$ component.
This result confirms that the $|-\rangle$ component condenses
before the $|0\rangle$ component
and surrounds the condensed $|+\rangle$ component.

\section{Conclusion}

We have studied the thermodynamics of Bose-Einstein
condensation for a gas of spin-1 atoms with
ferromagnetic and antiferromagnetic interactions using the mean field
Hartree-Fock-Popov theory and the semiclassical approximation
for the noncondensed components. Our results show that
for antiferromagnetic interactions, double phase transitions persist as
in a noninteracting gas: when $M>0$, first the $|+\rangle$ component
condenses, which is followed by the
condensation of the $|-\rangle$ component on further decreasing of the
temperature. The $|0\rangle$ component never condenses.
For ferromagnetic interactions, on the other hand, our calculations reveal
that the phase diagram becomes more complicated and a
triple condensation scenario arises with decreasing temperatures:
first the $|+\rangle$ component condenses,
which is followed by the second condensation of the $|-\rangle$ component,
and the third one for the $|0\rangle$ component. When the $|+\rangle$ and $|-\rangle$
components are the only condensed ones, they are immiscible. When all three
components condense and the temperature is lower than $T_4$, they become
miscible because of the presence of a large condensed $|0\rangle$ component.
We have compared the numerically computed
transition temperatures with that of an ideal gas as in Fig. \ref{pc}.
An overall lowering of the various transition temperatures due to
atom-atom interactions is seen, consistent with the case of a
single component interacting Bose gas, where the interaction-caused
shift to the transition temperature has been actively studied \cite{Holzmann01}.
Quite generally a repulsive
interaction tends to lower the transition temperature for a single component
Bose gas \cite{Str}. In the case of a spin-1 Bose gas considered here,
$c_0>0$ and $c_0\gg |c_2|$ constitutes an overall repulsive interaction.

Finally, we note there also exists the possibility of a ferromagnetic
phase transition for $^{87}$Rb atoms, in addition to the Bose-Einstein
condensation as studied here. In fact, as was investigated recently
by Gu and Klemm \cite{Gu03}, the ferromagnetic transition is
generally predicted to occur before, i.e. at temperatures higher
than, the Bose Einstein condensation. The present study, however,
remains unchanged because we treated the system within the
global constraint of the conservation of total magnetization,
distinct from that required for a separate ferromagnetic phase transition \cite{Gu03}.
As is evidenced from recent experiments, the total
magnetization $M$ is well conserved, even better than the conservation of the
total number $N$ \cite{Chapman03}.

\section*{ACKNOWLEDGMENTS}

This work is supported by NSF and NASA. We thank M.-S. Chang
for helpful discussions.

\end{document}